\documentclass[aps,prl,reprint,superscriptaddress,floatfix]{revtex4-2}

\usepackage{graphicx}
\usepackage{amsmath}
\usepackage{amssymb}
\usepackage{bm}
\usepackage{xcolor}
\graphicspath{{figure/}}
\newcommand{\rev}[1]{#1}
\newcommand{\dengrev}[1]{#1}
\newcommand{\dengrevxxi}[1]{#1}
\newcommand{\dengrevxxiv}[1]{#1}


\begin{document}

\title{Persistence of BKT phase transition in the 2D nonanalytic XY model}

\author{Sihan Hu}
\thanks{These authors contributed equally to this work.}
\affiliation{Hefei National Laboratory for Physical Sciences at the Microscale and Department of Modern Physics, University of Science and Technology of China, Hefei 230026, China}
\author{Xianzhi Pan}
\thanks{These authors contributed equally to this work.}
\affiliation{Hefei National Laboratory for Physical Sciences at the Microscale and Department of Modern Physics, University of Science and Technology of China, Hefei 230026, China}
\author{Kun Chen}
\email{chenkun@itp.ac.cn}
\affiliation{Institute of Theoretical Physics, Chinese Academy of Sciences, Beijing 100190, China}
\author{Yi Jiang}
\email{jiangyi@ustc.edu.cn}
\affiliation{Department of Modern Physics, University of Science and Technology of China, Hefei 230026, China}
\author{Youjin Deng}
\email{yjdeng@ustc.edu.cn}
\affiliation{Hefei National Laboratory for Physical Sciences at the Microscale and Department of Modern Physics, University of Science and Technology of China, Hefei 230026, China}
\affiliation{Hefei National Laboratory, University of Science and Technology of China, Hefei 230026, China}
\affiliation{College of Physics, Guizhou University, Guiyang 550025, China}

\date{\today}

\begin{abstract}
We study the two-dimensional XY model with the nonanalytic pair potential
$2[(1-\cos\delta)/2]^{p}$, whose small-angle law $\propto|\delta|^{2p}$
carries a cusp for $p<1$ and a flat bottom for $p>1$, invalidating the
harmonic spin-wave expansion. Two questions arise: the nature of the
low-temperature ($T$) phase and of the phase transition.
A naive energetic argument would predict genuine long-range order and an enhanced transition
temperature for $p<1$, and no transition at all for $p>1$. Large-scale Monte
Carlo simulations contradict both: for every $p>0$ the
\dengrevxxi{low-$T$ phase} is quasi-long-range ordered, with anomalous dimension
$\eta(T)\propto T^{1/p}$, and terminates at a Berezinskii--Kosterlitz--Thouless transition. 
Using a vortex-free noncompact lattice-field description and utilizing a duality transformation, 
we show that coarse-graining drives the height-difference distribution onto a
single Gaussian fixed point, renormalizing the cusp and flatness
into a finite harmonic stiffness that restores the spin-wave description and
the BKT scenario for all $p>0$.
\end{abstract}

\maketitle

\textit{Introduction.---}The two-dimensional (2D) XY model is a prototypical
system for topological phase transitions in a broad
variety of quantum and classical systems. 
With a continuous $\mathrm{O}(2)$ symmetry and short-range
interactions it cannot develop long-range order at any finite temperature $T$, as
dictated by the Mermin--Wagner theorem~\cite{mermin1966absence,hohenberg1967existence};
instead it undergoes the Berezinskii--Kosterlitz--Thouless (BKT)
topological transition $T_c$~\cite{berezinskii1971destruction,kosterlitz1973ordering,kosterlitz1974critical,jose1977renormalization}, driven by
vortex--antivortex unbinding. Below $T_c$ lies a quasi-long-range-ordered
(QLRO) phase with algebraically decaying correlations~\cite{nelson1977universal,kosterlitz2016kosterlitz}.
The paradigm has been
realized in thin $^4$He films~\cite{bishop1978study}, superconducting
arrays~\cite{resnick1981kosterlitz}, and 2D ultracold atomic
gases~\cite{hadzibabic2006berezinskii}. The microscopic phase-difference
energy, however, need not be the simple cosine of the textbook model. Modern
Josephson weak links and superconducting circuits can exhibit or engineer
higher harmonics in the current--phase relation~\cite{schrade2022protected,zhang2024second,willsch2024josephson}.
In driven polariton condensates, 
BKT diagnostics probe an effective stiffness rather than a fixed microscopic
quadratic coefficient~\cite{comaron2025coherence}. Programmable dipolar
XY simulators now also allow microscopic spin-wave dynamics to be measured
directly~\cite{chen2023continuous,chen2025spectroscopy}.

A recurring question is how robust \dengrevxxi{the QLRO phase and the BKT transition are} 
against modifications of the microscopic interaction~\cite{domany1984first,romano2002xy,vzukovivc2017xy,%
da2024first}. 
Let $H=\sum_{\langle ij\rangle}V(\delta_{ij})$ be the Hamiltonian, 
$\exp(-\beta H)$ be the Boltzmann weight ($\beta=1/T$), and
$\delta_{ij}=\theta_i-\theta_j$ be the angle difference between neighboring
spins. A paradigmatic example is the generalized potential introduced by Domany, Schick,
and Swendsen (DSS)
\begin{equation}
  V_{\mathrm{DSS}}(\delta)
  =-2\left(\frac{1+\cos\delta}{2}\right)^{1/p}.
  \label{eq:Hdss}
\end{equation}
\dengrevxxi{For $p=1$ it reduces to the standard XY model, while decreasing
$p$ makes the potential increasingly narrow without removing its quadratic
small-angle expansion. DSS reported that, for sufficiently narrow wells, the
QLRO phase can terminate at a first-order transition 
rather than at a BKT transition}~\cite{domany1984first}\dengrevxxi{---a scenario long}
debated~\cite{ota1994microcanonical,sinha2010finite} and later proven
rigorously for special cases by van Enter and
Shlosman~\cite{van2002first,enter2005provable}; analogous first-order behavior
was also reported in a nonlinear Heisenberg variant~\cite{blote2002phase}.

\begin{figure}
  \includegraphics[width=\columnwidth]{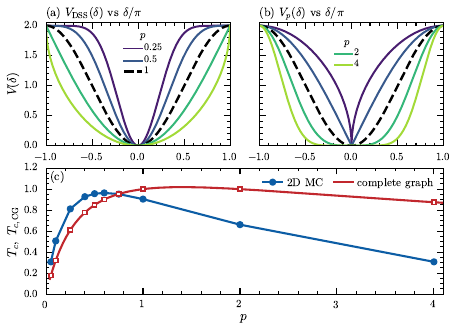}
  \caption{\label{fig:tc}%
    Pair potentials and transition temperatures. Panels (a) and (b) show,
    respectively, the DSS potential
    $V_{\mathrm{DSS}}(\delta)=2[1-\{(1+\cos\delta)/2\}^{1/p}]$ and the present
    nonanalytic XY potential
    $V_p(\delta)=2[(1-\cos\delta)/2]^p=2|\sin(\delta/2)|^{2p}$ for
    representative values of $p$.
    Filled circles in panel (c) show the BKT transition temperature $T_c$ of
    the 2D nonanalytic XY model.
    The red curve and open squares show the exact complete-graph transition
    $T_{c,\mathrm{CG}}$ in the same energy normalization. Both $T_c$ and
    $T_{c,\mathrm{CG}}$ vary nonmonotonically with $p$ and decrease toward the
    small-$p$ limit.}
\end{figure}

A more fundamental question is what happens when this quadratic
low-energy structure itself is removed.
Here we study a generalized XY model with 
\begin{equation}
  V_p(\delta)=2\left(\frac{1-\cos\delta}{2}\right)^{p}.
  \label{eq:Hintro}
\end{equation}
Both potentials yield the standard XY model at $p=1$.
Their wells are broader and flatter for $p>1$ and narrower for $p<1$
[Fig.~\ref{fig:tc}(a),(b)]. The DSS minimum remains quadratic for all $p>0$,
whereas $V_p(\delta)\sim2^{1-2p}|\delta|^{2p}$ at small angles.
Its curvature is therefore singular for $p<1$ and vanishes for $p>1$,
precluding the usual harmonic spin-wave expansion for $p\ne1$.
Related Mermin--Wagner questions were considered mathematically~\cite{shlosman1980phase,ioffe20022d},
but two important questions
remain: what is the nature of the low-$T$ phase and what is the fate of the BKT
transition?
\color{black}

\color{black}
A bare-energy estimate would suggest a very different picture. For $p<1$,
\dengrevxxi{the lowest nonzero twist mode} on a square lattice of linear size $L$ costs
$E_{\rm twist}\sim L^{2-2p}$, so the spin-wave
fluctuations appear increasingly suppressed with system size, seemingly
allowing true long-range order. 
Moreover, while
the entropy of a free vortex is $S_v\sim2\ln L$, its bare energy grows as
$L^{2-2p}$ for $p<1$, logarithmically at $p=1$, and remains finite for
$p>1$. Thus the usual energy--entropy argument would predict a transition
temperature that rises monotonically as $p$ is reduced below unity, and might
even disappear for $p>1$ where vortex proliferation is entropically favored.
\color{black}

\color{black}
However, our Monte Carlo (MC) simulations show that, for all $p$ studied,
the \dengrevxxiv{low-$T$ phase} remains \dengrevxxi{QLRO} and the transition
remains being BKT. The anomalous dimension $\eta$, 
governing the algebraic decay of two-point correlation function in low-$T$ phase,
varies continuously and approaches the
universal value $\eta(T_c)=1/4$ at the transition. 
On the other hand, unlike the linear relation $\eta \propto T$ for $p=1$, we have $\eta \propto T^{1/p}$. 
The phase boundary is nonmonotonic: \dengrevxxi{$T_c$} is strongly suppressed rather than enhanced
when $p$ is reduced below unity, and, in the $p>1$ regime, it also decreases as $p$ increases.
\color{black}

\color{black}
To reveal underlying mechanisms, we start with
the nonanalytic XY model on the complete graph, where sites are fully connected 
and, thus, spin waves and vortices are absent.
An angular Fourier decomposition of the spin distribution yields
orthogonal harmonic modes $m_n$ ($n=1,2,\ldots$), resulting in a Landau
mean-field theory of multiple angular, rather than spatial, modes.
The $n=1$ magnetic mode softens first, and
nonlinear entropy terms then couple this critical mode to the
still-massive higher harmonics, 
without generating additional transitions.
\dengrevxxi{The exact solution shows that the pair potential alone produces
nonmonotonic transition temperatures, paralleling the 2D phase boundary in
Fig.~\ref{fig:tc}(c).}

To capture low-$T$ spin-wave physics, 
we then consider the noncompact lattice field model on the square lattice 
\begin{equation}
  H_{\rm nc}=K\sum_{\langle ij\rangle}|h_i-h_j|^{2p},\qquad h_i\in\mathbb{R},
  \label{eq:Hnc}
\end{equation}
\dengrev{Real-space coarse graining shows that its strongly non-Gaussian
microscopic fluctuations flow toward Gaussian statistics.
\dengrevxxi{Moreover, the exact duality transformation shows that, due to the
curl-free constraint on each elementary plaquette, the local bond
weight becomes a characteristic function whose finite variance forces the
small-gradient action to begin quadratically for every $p>0$.}
Once this Gaussian infrared form is established, the homogeneity
of the noncompact model fixes the stiffness scaling and hence
$\eta(T)\propto T^{1/p}$. Restoring compactness then provides the basis for
the persistent BKT behavior.}
\color{black}

\begin{figure*}
  \includegraphics[width=\textwidth]{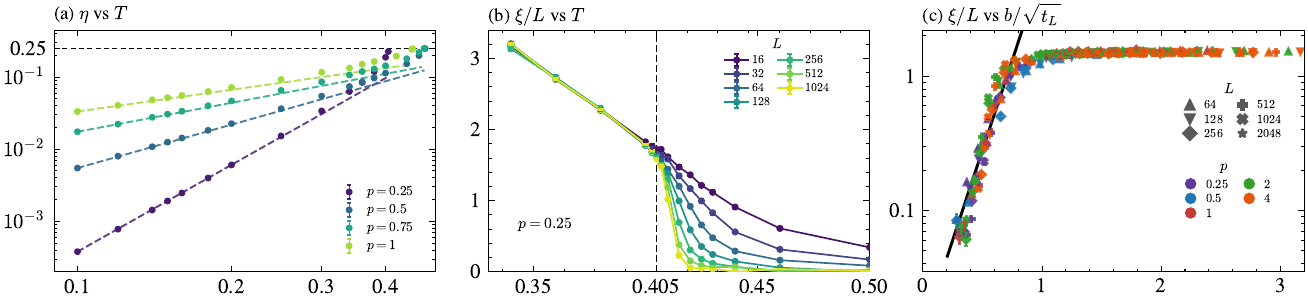}
  \caption{\label{fig:bkt}%
    BKT character of the 2D nonanalytic XY model.
    (a)~Anomalous dimension $\eta(T)$ in the \dengrevxxi{QLRO phase} for
    several exponents $p$, obtained from the reduced susceptibility
    $\chi_0\sim L^{2-\eta}$. On the
    log-log scale the low-$T$ data follow the power law
    $\eta\propto T^{1/p}$; the dashed lines mark the theoretical slope $1/p$
    (not fits). Toward $T_c$ each curve bends
    up to the universal value $\eta=1/4$ (black dashed), reflecting vortex
    renormalization beyond spin-wave theory.
    (b)~Correlation-length ratio $\xi/L$ versus $T$ near the
    transition at $p=0.25$ ($16\le L\le1024$): the curves for different $L$s cross
    near $T_c\simeq0.811$ (black dashed).
    (c)~BKT data collapse of $\xi/L$ against $b/\sqrt{t_L}$, with the rescaled
    variable $t_L=t[\ln L]^2$ and reduced temperature $t=(T-T_c)/T_c$
    ($T>T_c$). For each $p$, a nonuniversal metric factor $b$ is adjusted,
    with $b=(0.57,0.80,1.00,1.31,2.28)$ for
    $p=(0.25,0.5,1,2,4)$. The black line is for guiding eyes.}
\end{figure*}

\rev{{\color{black}\textit{Simulations and observables.---} We
perform large-scale simulations of Eq.~\eqref{eq:Hintro} for
$0.05\le p\le4$, up to $L=2048$, using \dengrevxxi{a combination of}
Swendsen--Wang cluster and event-chain Monte Carlo
algorithms~\cite{swendsen1987nonuniversal,michel2015event}. The
\dengrevxxiv{low-$T$ phase} is characterized through the magnetization
$m=N^{-1}\sum_j e^{i\theta_j}$ and the zero-momentum susceptibility
$\chi_0=N\langle|m|^2\rangle$. In a QLRO phase,
$\chi_0\sim L^{2-\eta}$, so \dengrevxxi{$\chi_0$ directly yields
the anomalous dimension $\eta(T)$}.
We also measure the second-moment correlation length,
\begin{align*}
  m_{\bf q}&=\frac{1}{N}\sum_j e^{i\theta_j}e^{i{\bf q}\cdot{\bf r}_j},
  &S({\bf q})&=N\langle|m_{\bf q}|^2\rangle,\\
  \xi&=\frac{1}{\sin(\pi/L)}
  \left[\frac{S({\bf 0})}{S({\bf q})}-1\right]^{1/2},
  &{\bf q}&=(2\pi/L,0).
\end{align*}
To locate the transition, we use the BKT condition $\eta(T_c)=1/4$, for which
$A_\chi\equiv\chi_0/L^{7/4}$ becomes asymptotically scale invariant up to
logarithmic corrections.  We fit all retained sizes to the corrected local
scaling form
\dengrev{$A_\chi(\ln L+C_1)^{-1/8}=F[(T-T_c)(\ln L+C_2)^2]$},
with $F$ being expanded quadratically near the transition.  Statistical errors are
obtained by parametric bootstrap, while \dengrevxxiv{the spread among stable
fits provides the method uncertainty}; see the Supplemental
Material (SM)~\cite{supplemental_material}.}}

\rev{{\color{black}\textit{Monte Carlo results.---}As shown in
Fig.~\ref{fig:bkt}(a), $\eta$ remains finite throughout the \dengrevxxiv{low-$T$}
phase for all $p$ studied, demonstrating quasi-long-range rather than true
long-range order. The simple power law is further revealed
\begin{equation*}
  \eta(T)\propto T^{1/p},
\end{equation*}
which generalizes the linear-in-$T$ spin-wave behavior for $p=1$. 
Figure~\ref{fig:bkt}(b) displays the typical BKT behavior in the
correlation-length ratio $\xi/L$: below $T_c$, the curves for different system
sizes collapse onto a common \dengrevxxiv{low-$T$ branch}, reflecting the critical
QLRO phase; above the transition, $\xi/L$ rapidly decreases with increasing
$L$.

The transition itself is consistent with the BKT universality class. The
extracted $\eta(T)$ approaches $1/4$ on approaching the transition, as seen in
Fig.~\ref{fig:bkt}(a). Above $T_c$, the correlation
length \dengrevxxi{takes the asymptotic form}
\begin{equation*}
\xi_\infty\sim \exp \left(b/\sqrt{t} \right), \qquad t= (T-T_c)/T_c .
\end{equation*}
For $\xi \sim {\rm O}(L)$, finite-size scaling theory gives 
the scaling combination $b/\sqrt{t_L}$, with $t_L=t(\ln L)^2$. Using the $T_c$ determined
from $A_\chi$, the data for different $L$ and $p$ collapse onto
common scaling curves when plotted against this variable, as shown in
Fig.~\ref{fig:bkt}(c). Together, the approach of $\eta$ to the universal value
$1/4$ and the characteristic essential-singularity scaling establish that the
transition remains in the BKT universality class throughout the studied range
of $p$. 
}}

\begin{figure*}[!t]
  \includegraphics[width=\textwidth]{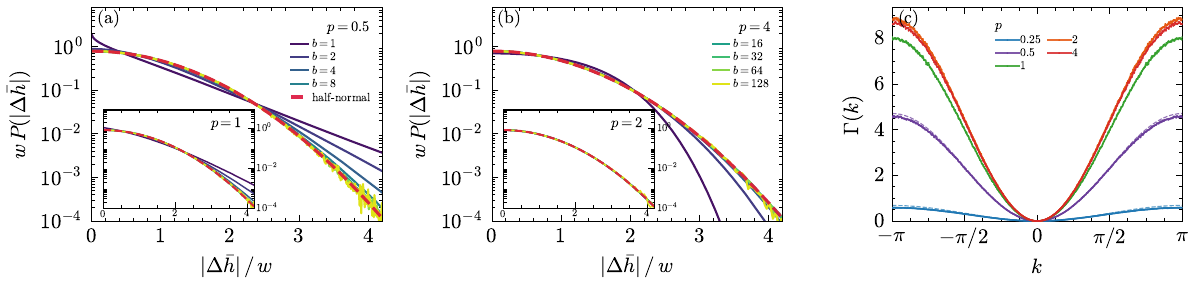}
  \caption{\label{fig:nc_cg}\color{black}%
    Gaussianization in the \dengrevxxiv{2D} noncompact lattice field
    model.
    (a),(b)~Real-space coarse-graining at $L=1024$. At each block
    scale $b$, nearest-neighbor height differences are standardized
    by their width
    $w^2=\langle(\Delta\bar h)^2\rangle$, giving the density
    $wP(|\Delta\bar h|)$ of
    $u=|\Delta\bar h|/w$. Curves are colored by block size from $b=1$ to $b=128$; 
    the red dashed line is for the standard Gaussian distribution. 
    For $p<1$, the distribution approaches a Gaussian from a sharper,
    heavy-tailed form, whereas for $p>1$ it approaches from a flatter,
    light-tailed form; the $p=1$ case is Gaussian already at $b=1$.
    (c)~Static inverse propagator $\Gamma(k)\equiv S(k)^{-1}$ at $L=512$.
    The dashed lines show the corresponding Gaussian lattice forms, with
    amplitudes determined by the long-wavelength stiffness.}
\end{figure*}

\rev{{\color{black}\textit{Complete-graph analysis.---}
\dengrevxxi{The nonmonotonic $T_c(p)$ raises a natural question: how much of
this behavior is encoded in the nonlinear pair potential before spatial
fluctuations and vortices are involved?}
\dengrevxxiv{We then consider the fully connected model},
\begin{equation}
H_{\mathrm{CG}}
=
\frac{zJ}{N}
\sum_{i<j}
V_p(\theta_i-\theta_j),
\label{eq:cg_hamiltonian}
\end{equation}
\dengrevxxiv{where $z$ represents the coordination number of the original lattice and
$J$ is the bond coupling.}
In the thermodynamic ($N\to \infty$) limit, a configuration \dengrevxxiv{on the complete graph} is described
by its normalized angular distribution $\rho(\theta)$.
\dengrevxxiv{The} free energy contains the mean pair-interaction energy
together with the entropic contribution
$T\int d\theta\,\rho(\theta)\ln\rho(\theta)$.
At sufficiently high temperature, the latter stabilizes
the uniform disordered state,
$\rho_0(\theta)=1/(2\pi)$.

Because $V_p$ depends only on the angular difference, the
interaction energy is diagonal in angular Fourier modes.
Let $v_n(p)$ denote the Fourier coefficients of the pair
potential and
$m_n=\int d\theta\,\rho(\theta)e^{in\theta}$
the corresponding harmonics of $\rho(\theta)$ (see SM for details).
Here $m_1$ is the ordinary magnetization, while
\dengrev{$m_{n>1}$ describe higher angular harmonics.}
Expanding the free energy, for small deviations from
$\rho_0$, yields
\begin{equation}
\Delta f^{(2)}
=
\sum_{n\geq1}
\bigl[T+zJ v_n(p)\bigr]|m_n|^2.
\label{eq:cg_quadratic_body}
\end{equation}
This gives a multimode Landau description of possible instabilities of the
disordered state.

For the present potential, $v_1$ is the uniquely most
negative Fourier coefficient.
The uniform distribution first loses stability
in the $n=1$ magnetic channel at
\begin{equation}
T_{c,\mathrm{CG}}
=
-zJ v_1(p)
=
2zJ\,
\frac{\Gamma\!\left(p+\tfrac12\right)}
{\sqrt{\pi}(p+1)\Gamma(p)}.
\label{eq:cg_tc}
\end{equation}
\dengrevxxiv{The higher-order modes $m_{n>1}$ remain noncritical at $T_{c,\mathrm{CG}}$
and contribute only to corrections in the effective Landau free
energy for $m_1$. The critical temperature grows linearly for small $p$,
$T_{c,\mathrm{CG}}\sim 2zJ\,p$,  reaches its maximum at
$p\simeq1.41$, and decreases as $T_{c,\mathrm{CG}}\sim
2zJ/\sqrt{\pi p}$ for $p\to\infty$ (Fig.~\ref{fig:tc}(c)).}}}

\rev{{\color{black}\textit{Duality and emergent Gaussian elasticity.---}
The anomalous \dengrevxxiv{low-$T$ law} in Fig.~\ref{fig:bkt}(a) calls for a
direct account of the low-temperature spin-wave expansion.
We therefore return to the
\dengrevxxiv{noncompact lattice field model} in Eq.~\eqref{eq:Hnc}, thereby
preserving the full nonlinear
bond weight $e^{-g|\Delta_\mu h|^{2p}}$, with $g=K/T$. Taking
$h_i\in\mathbb R$ lifts compactness and hence removes vortices, isolating
the nontopological fluctuations of the \dengrevxxiv{low-$T$ phase}.
\dengrevxxi{We expose its infrared structure through an exact lattice
duality.}

Introducing the bond variables $u_{i\mu}=\Delta_\mu h_i$, where
$\Delta_\mu$ denotes a forward lattice difference, makes the nonlinear
energy local on each bond. Because these variables originate from a
single-valued height field, their circulation around every plaquette must
vanish. A real Lagrange multiplier $\phi_i$ on each plaquette enforces this
compatibility condition. Integrating out the bond variables then leaves
$\phi$ as a scalar field on the dual square lattice, with the exact
representation \dengrevxxiv{(see SM for details)}
\begin{equation}
Z_g\propto
\int\mathcal D\phi\,
\prod_i W(\Delta_x\phi_i)W(\Delta_y\phi_i).
\label{eq:dual_rep}
\end{equation}
Here \dengrev{$W(a)=\int_{-\infty}^{\infty}du\,
e^{-g|u|^{2p}+iau}$} is the Fourier transform of the single-bond thermal
weight; \dengrev{$W(0)$ corresponds to a locally flat dual field, and
$W(a)/W(0)$ is the normalized function of the bond
fluctuations.}
At long wavelengths its argument becomes small, and this
distribution has a finite variance $\kappa_2=\langle u^2\rangle_g$ for
every $p>0$. Its local logarithm consequently begins as
\begin{equation}
-\ln\frac{W(a)}{W(0)}
=
\frac{\kappa_2}{2}a^2+o(a^2),
\qquad a\to0 .
\label{eq:dual_cumulant}
\end{equation}
The emergent quadratic response thus reflects thermal smearing over the
full bond distribution: it is controlled by its finite variance rather
than by the bare curvature of $|u|^{2p}$, which may diverge or vanish at
the origin. The leading dual action is therefore quadratic, while higher
local gradient powers are irrelevant near the \dengrevxxiv{2D} Gaussian
fixed line. Inverting the same Gaussian duality gives Gaussian infrared
elasticity for the original height field.

Simulations are also performed for the lattice-field model by Eq.~(\ref{eq:Hnc}),
and the real-space coarse graining in Fig.~\ref{fig:nc_cg} shows directly how
these very different microscopic bond distributions approach the same
Gaussian form. After normalization to the \dengrevxxiv{corresponding variance}, the $p<1$
distributions are initially more sharply peaked at the origin and have
heavier tails than a Gaussian, whereas the $p>1$ distributions are flatter
near the center and have more strongly suppressed tails; $p=1$ is Gaussian
already at the microscopic scale. Repeated block averaging progressively
erases both forms of microscopic non-Gaussianity. The inverse
propagator in Fig.~\ref{fig:nc_cg}(c) is correspondingly quadratic at long
wavelengths and characterizes the low-temperature spin-wave fluctuations of
the model in Eq.~\eqref{eq:Hintro}. For $p<1$, the coarse-grained dispersion
is flatter despite the sharp bare potential, indicating that its large local
energy cost does not translate into a larger long-wavelength stiffness. The
resulting softer spin-wave fluctuations enhance thermal angular fluctuations,
consistent with the suppression of $T_c$. For $p>1$, the noncompact Gaussian
field does not include the vortex-core contribution present in the compact
nonanalytic XY model. The reduction of the vortex-core energy then becomes
increasingly important for the continued decrease of $T_c$.

\dengrev{With the infrared Gaussian form now established, the temperature
dependence in Fig.~\ref{fig:bkt}(a) follows from the homogeneity of the
noncompact Hamiltonian. Under the explicit field rescaling
$h_{\boldsymbol r}=g^{-1/(2p)}\widetilde h_{\boldsymbol r}$, the measure at
coupling $g$ is mapped onto that at $g=1$. Consequently,
\begin{equation}
\left\langle[h(\boldsymbol r)-h(\boldsymbol 0)]^2\right\rangle_{g,p}
=g^{-1/p}
\left\langle[\widetilde h(\boldsymbol r)-\widetilde h(\boldsymbol 0)]^2
\right\rangle_{1,p}.
\label{eq:body_height_rescaling}
\end{equation}
The rescaled field $\widetilde h$ on the right obeys the conventional BKT
result
$\langle[\widetilde h(\boldsymbol r)-\widetilde h(\boldsymbol 0)]^2\rangle_{1,p}
=(\pi K_R(1,p))^{-1}\ln r+O(1)$, which directly gives
$K_R(g,p)=g^{1/p}K_R(1,p)$ and hence
$\eta=1/(2\pi K_R)\propto T^{1/p}$.}


Finally, restoring compactness reintroduces vortices into this emergent
harmonic medium. Their far-field energy is then logarithmic rather than
being governed by the bare $|\nabla\theta|^{2p}$ power law, recovering the
elastic basis of the BKT mechanism. The microscopic nonlinearity still
affects the renormalized stiffness, vortex-core physics, and crossover
scales.}}

\rev{{\color{black}\textit{Discussion.---}In summary, our numerical and analytical results reveal a unified
infrared picture for the nonanalytic XY model. Although bare twist
and vortex energies suggest qualitatively different long-distance
behavior on the two sides of $p=1$, the \dengrevxxiv{low-$T$ phase} remains
\dengrevxxi{QLRO} and the transition remains of BKT type for
every finite $p>0$. The microscopic nonlinearity nevertheless leaves
clear signatures: $\eta(T)\propto T^{1/p}$ at \dengrevxxiv{low $T$}, while
$T_c(p)$ is finite but nonmonotonic. The complete-graph theory
identifies the angular energy--entropy competition that controls the
ordering scale, whereas the \dengrevxxiv{2D dual-field analysis} explains why the
spatially fluctuating system ultimately develops Gaussian infrared
elasticity.

The failure of bare-energy reasoning is especially clear at small $p$.
The bond potential resembles an almost flat golf course with a deep,
vanishingly narrow hole. The hole offers a substantial energy gain, but
only a vanishing fraction of angular configurations lies within it,
so the gain is offset by
the loss of configurational entropy. The complete-graph calculation
makes this competition explicit through the full angular spectrum of
the interaction. The same lesson applies to
vortices in two dimensions. The bare energy of a vortex embedded in a
zero-temperature background is not its free-energy cost in the
fluctuating system; the nontopological fluctuations must first be
integrated out to determine the elastic medium in which vortices
interact.

\dengrevxxiv{Unlike long-range XY/$\mathrm{O}(n)$ models, where interactions
$\mathcal J(r)\sim r^{-(2+\sigma)}$ generate the nonlocal kernel
$|\mathbf{k}|^\sigma$---equivalently, a fractional Laplacian
$(-\nabla^2)^{\sigma/2}$---and can support true long-range order in
2D for $\sigma<2$~\cite{fisher1972critical,xiao2025two,yao2025nonclassical,li2026epsilon},
the nonanalyticity here is confined to the local bond weight, and, moreover, 
the associated bond field remains curl free on each elementary plaquette.
The broader coarse-graining
lesson is relevant to granular and ultrasoft matter, where
nonanalytic power laws arise naturally, while the BKT conclusion is most
directly testable in Josephson-junction arrays, whose current--phase relations
can contain substantial higher harmonics. In such phase-coupled systems,
strong microscopic nonlinearity need not change the infrared universality
class; its clearest trace may instead be the anomalous low-$T$ law
($\eta(T)\propto T^{1/p}$). The microscopic harmonic theory may be absent, yet
harmonic elasticity---and with it the BKT mechanism---can be rebuilt
collectively in the infrared.}}}

\begin{acknowledgments}
S.H., X.P., and Y.D. are supported by the National Natural Science Foundation
of China under Grant No.~12275263, the Quantum Science and
Technology---National Science and Technology Major Project under Grant
No.~2021ZD0301900, and the Natural Science Foundation of Fujian Province of
China under Grant No.~2023J02032. K.C. is supported by the Strategic Priority
Research Program of the Chinese Academy of Sciences under Grant
No.~XDB1680102, the National Key Research and Development Program of China
under Grant No.~2024YFA1408604, and the National Natural Science Foundation of
China under Grants No.~12474245 and No.~12447103.
\end{acknowledgments}

\bibliography{refs}

\clearpage
\onecolumngrid
\setcounter{equation}{0}
\renewcommand{\theequation}{S\arabic{equation}}
\setcounter{figure}{0}
\renewcommand{\thefigure}{S\arabic{figure}}
\setcounter{table}{0}
\renewcommand{\thetable}{S\arabic{table}}
\setcounter{secnumdepth}{2}
\renewcommand{\thesection}{\Roman{section}}
\renewcommand{\thesubsection}{\Alph{subsection}}
\begin{center}
  \textbf{Supplemental Material}
\end{center}
\section{Overview of this Supplemental Material}

\dengrevxxiv{This Supplemental Material is organized into three parts.
Section~II derives the complete-graph mean-field theory and its angular
ordering channels. Section~III gives the exact bond duality and the emergence
of Gaussian infrared elasticity in the noncompact lattice field model.
Section~IV summarizes the Monte Carlo observables, the global finite-size
scaling analysis, and the transition-temperature estimates used in the main
text.}

\section{\dengrev{Complete-graph mean-field theory}}%
\label{sec:cg}

\subsection{Model and angular Fourier decomposition}

We consider the fully connected model
\begin{equation}
  H_{\mathrm{CG}}
  =
  \frac{zJ}{N}
  \sum_{1\leq i<j\leq N}
  V_p(\theta_i-\theta_j),
  \qquad
  V_p(\varphi)
  =
  2\left|\sin\frac{\varphi}{2}\right|^{2p}.
  \label{eq:Hcg}
\end{equation}
Here $z$ is the coordination number of the original lattice and $J$
is its nearest-neighbor coupling.
The factor $1/N$ keeps the energy per spin finite in the thermodynamic
limit. A macroscopic state is described by its angular distribution
$\rho(\theta)$, normalized as
\begin{equation}
  \int_0^{2\pi}d\theta\,\rho(\theta)=1.
\end{equation}
The corresponding free-energy density is
\begin{equation}
  f[\rho]=
  \frac{zJ}{2}
  \int_0^{2\pi}\!d\theta
  \int_0^{2\pi}\!d\phi\,
  \rho(\theta)\rho(\phi)
  V_p(\theta-\phi)
  +T\int_0^{2\pi}\!d\theta\,
  \rho(\theta)\ln\rho(\theta).
\label{eq:fc_free_energy}
\end{equation}
At high temperature, the stationary solution is the uniform distribution
\begin{equation}
  \rho_0(\theta)=\frac{1}{2\pi}.
\end{equation}

Because the interaction depends only on the angular difference, we
introduce the Fourier coefficients
\begin{equation}
  V_p(\varphi)
  =
  \sum_{n=-\infty}^{\infty}
  v_n e^{-in\varphi},
  \qquad
  m_n
  =
  \int_0^{2\pi}d\theta\,
  \rho(\theta)e^{in\theta},
  \label{eq:cg_fourier_def}
\end{equation}
where $m_0=1$, $m_{-n}=m_n^*$, and $v_{-n}=v_n$. The quantities
$m_n$ are harmonics of the single-spin angular distribution; in
particular, $m_1$ is the usual complex magnetization. The interaction
energy becomes
\begin{equation}
  e[\rho]
  =
  \dengrev{\frac{zJ}{2}v_0
  +zJ\sum_{n=1}^{\infty}v_n|m_n|^2}.
  \label{eq:cg_fourier_energy}
\end{equation}

For
\begin{equation}
  V_p(\varphi)
  =
  2\left|\sin\frac{\varphi}{2}\right|^{2p},
\end{equation}
the Fourier coefficients are
\begin{equation}
  v_n=(-1)^n\frac{\Gamma(2p+1)}{2^{2p-1}}
  \left[\Gamma(p+n+1)\Gamma(p-n+1)\right]^{-1}.
\label{eq:vp_fourier}
\end{equation}
For integer $p$, the Fourier series terminates at $|n|=p$, whereas
for noninteger $p$ it contains infinitely many angular harmonics.

\subsection{Instability of the uniform distribution}

To expand about the uniform distribution, we write
\begin{equation}
  \rho(\theta)
  =
  \frac{1}{2\pi}
  \left[1+\eta(\theta)\right],
  \qquad
  \eta(\theta)
  =
  \sum_{n\neq0}m_ne^{-in\theta}.
  \label{eq:rho_eta}
\end{equation}
Using
\begin{equation}
  (1+\eta)\ln(1+\eta)
  =
  \eta+\frac{\eta^2}{2}
  -\frac{\eta^3}{6}
  +\frac{\eta^4}{12}
  +O(\eta^5),
  \label{eq:entropy_series}
\end{equation}
the quadratic free energy is
\begin{equation}
  \Delta f^{(2)}
  =\frac{1}{2}\sum_{n\neq0}
  \left[T+zJ v_n(p)\right]|m_n|^2
  =\sum_{n\geq1}
  \left[T+zJ v_n(p)\right]|m_n|^2.
\label{eq:cg_quadratic}
\end{equation}

This quadratic form defines a spectrum of angular ordering channels.
The entropy contributes the same quadratic cost $T$ to every harmonic,
while the interaction shifts the $n$th channel by $zJ v_n(p)$. A channel
with $v_n<0$ becomes unstable at
\begin{equation}
  T_n=-zJ v_n.
  \label{eq:candidate_instability}
\end{equation}

The Fourier coefficients obey
\begin{equation}
  \frac{v_n}{v_{n-1}}
  =
  -\frac{p-n+1}{p+n},
  \label{eq:vn_recurrence}
\end{equation}
which implies that the nonzero coefficients decrease in magnitude with
$n$. Since $v_1<0$, the $n=1$ channel has the highest candidate
instability temperature. The uniform distribution therefore first
becomes unstable in the magnetic channel, giving
\begin{equation}
  T_{c,\mathrm{CG}}
  =
  -zJ v_1
  =
  2zJ\,
  \frac{\Gamma\!\left(p+\frac12\right)}
  {\sqrt{\pi}(p+1)\Gamma(p)}.
  \label{eq:Tc_cg}
\end{equation}

Its limiting forms are
\begin{equation}
  \frac{T_{c,\mathrm{CG}}}{zJ}
  =
  2p\left[
    1-(1+2\ln2)p+O(p^2)
  \right],
  \qquad p\rightarrow0,
  \label{eq:Tc_small_p}
\end{equation}
and
\begin{equation}
  \frac{T_{c,\mathrm{CG}}}{zJ}
  =
  \frac{2}{\sqrt{\pi p}}
  \left[
    1-\frac{9}{8p}+O(p^{-2})
  \right],
  \qquad p\rightarrow\infty.
\label{eq:Tc_large_p}
\end{equation}

\subsection{Higher-order mode coupling}

At quadratic order, the angular harmonics form independent candidate
ordering channels. Beyond quadratic order, the entropy couples them:
the $n=1$ magnetic mode is the primary instability, while the higher
harmonics are generated successively as $m_n=O(m_1^n)$ and modify the
higher-order Landau coefficients. To determine the free energy through
order $|m_1|^4$, it is sufficient to retain the second harmonic:
\begin{equation}
\begin{aligned}
  \Delta f
  ={}&
  \left[T+zJ v_1(p)\right]|m_1|^2
  +
  \left[T+zJ v_2(p)\right]|m_2|^2
  \\
  &-
  \frac{T}{2}
  \left(
    m_1^2m_2^*
    +
    m_1^{*2}m_2
  \right)
  +
  \frac{T}{2}|m_1|^4
  +
  O(|m_1|^6).
\end{aligned}
\label{eq:cg_landau_complex}
\end{equation}

Minimizing over $m_2$ gives
\begin{equation}
  m_2
  =
  \frac{T}
  {2\left[T+zJ v_2(p)\right]}
  m_1^2
  +
  O(m_1^4),
  \label{eq:m2_slaving}
\end{equation}
and the resulting effective free energy is
\begin{equation}
  \Delta f_{\mathrm{eff}}
  =
  \left[T+zJ v_1(p)\right]|m_1|^2
  +
  u_{\mathrm{eff}}(T)|m_1|^4
  +
  O(|m_1|^6),
  \label{eq:cg_effective_phi4}
\end{equation}
with
\begin{equation}
  u_{\mathrm{eff}}(T)
  =
  \frac{T}{2}
  -
  \frac{T^2}
  {4\left[T+zJ v_2(p)\right]}.
  \label{eq:cg_ueff}
\end{equation}
Using $v_2/v_1=-(p-1)/(p+2)$, one obtains
\begin{equation}
  u_{\mathrm{eff}}(T_{c,\mathrm{CG}})
  =
  \frac{3p}{4(2p+1)}
  T_{c,\mathrm{CG}}
  >0.
  \label{eq:cg_quartic}
\end{equation}
Hence
\begin{equation}
  |m_1|^2
  =
  \frac{T_{c,\mathrm{CG}}-T}
  {2u_{\mathrm{eff}}(T_{c,\mathrm{CG}})}
\end{equation}
near the instability.

The complete graph therefore exhibits competition among angular
ordering channels at quadratic order, followed by nonlinear coupling
to the noncritical higher harmonics. Only the $n=1$ magnetic mode
becomes critical, while the higher harmonics shape the ordered angular
distribution and renormalize its higher-order Landau coefficients.

\section{\dengrev{Duality and Gaussianization in the noncompact lattice field model}}

The smooth sector associated with the nonlinear interaction is described
by the noncompact lattice field model
\begin{equation}
Z_g
=
\int \prod_r dh_r\,
\exp\left[
-g\sum_{r,\mu}|\Delta_\mu h_r|^{2p}
\right],
\qquad
g\equiv\frac{K}{T},
\label{appB:partition}
\end{equation}
where
\(
\Delta_\mu h_r=h_{r+\hat\mu}-h_r
\)
and the uniform height mode is omitted.

The essential point is that the nonanalyticity resides in the local bond
measure, rather than in a nonlocal infrared kernel. The derivation below
shows how collective fluctuations of this local measure generate a
quadratic long-wavelength response even when the microscopic curvature at
$\Delta_\mu h=0$ is singular or vanishing.

\subsection{Exact bond duality}

The dual transformation is most transparent after replacing the
heights by bond gradients,
\begin{equation}
u_x(r)=h_{r+\hat x}-h_r,
\qquad
u_y(r)=h_{r+\hat y}-h_r.
\label{appB:bond-gradients}
\end{equation}
Since these bond variables originate from a single-valued height
field, their oriented sum around every plaquette vanishes,
\begin{equation}
C_r
\equiv
\Delta_yu_x(r)-\Delta_xu_y(r)
=
0.
\label{appB:curl-constraint}
\end{equation}
The partition function can therefore be written as
\begin{equation}
Z_g
\propto
\int\mathcal D u_x\,\mathcal D u_y
\prod_r\delta(C_r)\,
\exp\left[
-g\sum_r
\left(
|u_x(r)|^{2p}+|u_y(r)|^{2p}
\right)
\right].
\label{appB:constrained-partition}
\end{equation}

Introducing a real field $\phi_r$ to impose the plaquette constraint,
\begin{equation}
\prod_r\delta(C_r)
\propto
\int\mathcal D\phi\,
\exp\left(
i\sum_r\phi_rC_r
\right),
\label{appB:constraint-field}
\end{equation}
gives
\begin{equation}
Z_g
\propto
\int
\mathcal D u_x\,\mathcal D u_y\,\mathcal D\phi\,
\exp\left[
-g\sum_{r,\mu}|u_\mu(r)|^{2p}
+i\sum_r\phi_rC_r
\right].
\label{appB:joint-partition}
\end{equation}
A discrete integration by parts yields
\begin{equation}
\sum_r\phi_rC_r
=
\sum_r
\left[
-(\Delta_y\phi_r)u_x(r)
+
(\Delta_x\phi_r)u_y(r)
\right].
\label{appB:integration-parts}
\end{equation}

At fixed $\phi$, each bond variable now appears only in its own local
term. The integrations over the bond variables therefore reduce to a
product of one-dimensional integrals:
\begin{align}
Z_g\propto
\int\mathcal D\phi\prod_r
&\left[
\int_{-\infty}^{\infty}du_x\,
\exp\left(
-g|u_x|^{2p}
-i(\Delta_y\phi_r)u_x
\right)
\right]
\nonumber\\
{}\times&
\left[
\int_{-\infty}^{\infty}du_y\,
\exp\left(
-g|u_y|^{2p}
+i(\Delta_x\phi_r)u_y
\right)
\right].
\label{appB:factorized-integrals}
\end{align}
Defining the common one-bond integral
\begin{equation}
W(a)
=
\int_{-\infty}^{\infty}du\,
\exp\left(
-g|u|^{2p}+iau
\right),
\label{appB:bond-transform}
\end{equation}
and using $W(-a)=W(a)$, we obtain the product representation
\begin{equation}
Z_g
\propto
\int\mathcal D\phi
\prod_r
W(\Delta_x\phi_r)\,
W(\Delta_y\phi_r).
\label{appB:product-representation}
\end{equation}

For a finite periodic lattice, this expression is understood in a fixed
zero-winding sector, with the uniform height mode removed. The other two
harmonic winding sectors contribute only global finite-size modes and do not
modify the local infrared response considered below.

Thus $\phi$, initially introduced only to impose the compatibility
constraint, becomes a scalar field on the dual lattice. Moreover,
$W(a)/W(0)$ is the characteristic function of the thermal single-bond
distribution. This identification turns the long-wavelength expansion of
the dual theory into a statement about the cumulants of the full bond
distribution.

Since $W(0)>0$ and $W(a)$ is continuous, it remains positive for
sufficiently small $|a|$. In this neighborhood we define
\begin{equation}
\Psi(a)
\equiv
-\ln\frac{W(a)}{W(0)},
\qquad
\Psi(0)=0,
\label{appB:dual-potential}
\end{equation}
so that
\begin{equation}
Z_g
\propto
\int\mathcal D\phi\,
e^{-S_{\mathrm{dual}}[\phi]},
\label{appB:dual-partition}
\end{equation}
with
\begin{equation}
S_{\mathrm{dual}}[\phi]
=
\sum_r
\left[
\Psi(\Delta_x\phi_r)
+
\Psi(\Delta_y\phi_r)
\right].
\label{appB:dual-action}
\end{equation}

\subsection{Gaussian infrared limit}

The small-gradient expansion is determined by the cumulants of the
normalized single-bond measure
\begin{equation}
P_g(u)
=
\frac{
e^{-g|u|^{2p}}
}{
\displaystyle
\int_{-\infty}^{\infty}du\,
e^{-g|u|^{2p}}
}.
\label{appB:bond-measure}
\end{equation}
\dengrev{By definition,
\begin{equation}
\frac{W(a)}{W(0)}
=\int_{-\infty}^{\infty}du\,P_g(u)e^{iau}
=\left\langle e^{iau}\right\rangle_{P_g},
\qquad
\Psi(a)=-\ln\left\langle e^{iau}\right\rangle_{P_g}.
\label{appB:characteristic-function}
\end{equation}}
Because $P_g(u)$ is even, only even cumulants appear:
\begin{equation}
\Psi(a)
=
\frac{\kappa_2}{2}a^2
-
\frac{\kappa_4}{24}a^4
+
o(a^4),
\label{appB:cumulant-expansion}
\end{equation}
where
\begin{equation}
\kappa_2
=
\langle u^2\rangle_g
=
g^{-1/p}
\frac{
\Gamma\left(\frac{3}{2p}\right)
}{
\Gamma\left(\frac{1}{2p}\right)
}.
\label{appB:second-cumulant}
\end{equation}
All polynomial moments are finite for every $p>0$, so the expansion
is well defined to any fixed finite order. For $p<1/2$, the
corresponding infinite Taylor series need not have a nonzero radius
of convergence.

The finite variance is the crucial fact. For $p<1$ the bare curvature
of $|u|^{2p}$ at the origin is singular, while for $p>1$ it vanishes,
but neither behavior controls the long-wavelength expansion. Instead,
the quadratic response is generated by thermal smearing over the full
bond distribution and is fixed at leading order by $\kappa_2$.

The leading term of the dual action is therefore Gaussian,
\begin{equation}
S_{\mathrm{dual}}^{(2)}
=
\frac{\kappa_2}{2}
\int d^2x\,(\nabla\phi)^2.
\label{appB:gaussian-dual}
\end{equation}
Higher cumulants generate operators containing four or more first
derivatives. Along the two-dimensional Gaussian fixed line, $\phi$
is dimensionless, so an operator containing $2n$ first derivatives
has scaling dimension $2n$. Its coupling has RG eigenvalue
\begin{equation}
y_{2n}=2-2n,
\label{appB:rg-eigenvalue}
\end{equation}
which is negative for every $n>1$. The higher-gradient terms are
therefore irrelevant, leaving a Gaussian long-wavelength theory
with a finite quadratic coefficient.

At $p=1$, the transformation is Gaussian at every scale:
\begin{equation}
W(a)
=
\sqrt{\frac{\pi}{g}}\,
e^{-a^2/(4g)},
\qquad
\Psi(a)=\frac{a^2}{4g}.
\label{appB:harmonic-check}
\end{equation}
For $p\neq1$, the higher cumulants control the approach to the
Gaussian fixed line. Although their associated operators are irrelevant,
they can produce a finite renormalization of the quadratic coefficient
before the flow reaches the fixed line. Accordingly, $\kappa_2$ is only
the bare quadratic coefficient in the dual representation and should not
be identified directly with the fully renormalized stiffness of the
original height field.

Figure~3 of the main text displays this flow at two distinct levels. After
normalization to the same variance, the $p<1$ bond distributions approach
a Gaussian from a sharper, heavy-tailed form, whereas the $p>1$
distributions approach from a flatter form with more strongly suppressed
tails; the $p=1$ distribution is Gaussian already at the microscopic
scale. Repeated block averaging progressively erases both kinds of
microscopic non-Gaussianity. This Gaussianization of a one-bond marginal
is not by itself sufficient to establish a Gaussian field theory. The
simultaneous observation $S(k_x)\propto k_x^{-2}$ at small $k_x$ supplies
the complementary field-level evidence for Gaussian infrared elasticity.

\subsection{Stiffness scaling}

A quadratic dual action corresponds to a quadratic long-wavelength
gradient theory for the original height field. We write its infrared
action as
\begin{equation}
S_{\mathrm{IR}}[h]
=
\frac{K_R(g,p)}{2}
\int d^2x\,(\nabla h)^2,
\label{appB:ir-height-action}
\end{equation}
where $K_R$ is the renormalized long-wavelength stiffness. With
distances measured in lattice units, the height fluctuations obey
\begin{equation}
\left\langle
[h(\boldsymbol r)-h(\boldsymbol 0)]^2
\right\rangle_{g,p}
=
\frac{1}{\pi K_R(g,p)}\ln r
+
O(1).
\label{appB:height-logarithm}
\end{equation}
The dependence on $g$ follows directly from the homogeneity of the
microscopic action. Under
\begin{equation}
h_r
=
g^{-1/(2p)}\widetilde h_r,
\label{appB:height-rescaling}
\end{equation}
one has
\begin{equation}
g\sum_{r,\mu}|\Delta_\mu h_r|^{2p}
=
\sum_{r,\mu}|\Delta_\mu\widetilde h_r|^{2p}.
\label{appB:rescaled-action}
\end{equation}
The measure for $\widetilde h$ is therefore the $g=1$ measure, and
\begin{align}
\left\langle
[h(\boldsymbol r)-h(\boldsymbol 0)]^2
\right\rangle_{g,p}
&=
g^{-1/p}
\left\langle
[\widetilde h(\boldsymbol r)
-\widetilde h(\boldsymbol 0)]^2
\right\rangle_{1,p}
\nonumber\\
&=
\frac{g^{-1/p}}{\pi K_R(1,p)}\ln r
+
O(1).
\label{appB:rescaled-correlator}
\end{align}
Comparison with Eq.~\eqref{appB:height-logarithm} gives
\begin{equation}
K_R(g,p)
=
g^{1/p}K_R(1,p).
\label{appB:stiffness-scaling}
\end{equation}

For a Gaussian field,
\begin{equation}
\left\langle
e^{i[h(\boldsymbol r)-h(\boldsymbol 0)]}
\right\rangle
=
\exp\left[
-\frac{1}{2}
\left\langle
[h(\boldsymbol r)-h(\boldsymbol 0)]^2
\right\rangle
\right].
\label{appB:gaussian-vertex}
\end{equation}
Combining Eqs.~\eqref{appB:height-logarithm} and
\eqref{appB:gaussian-vertex} gives
\begin{equation}
\left\langle
e^{ih(\boldsymbol r)}
e^{-ih(\boldsymbol 0)}
\right\rangle
\sim
r^{-\eta},
\qquad
\eta(g,p)=\frac{1}{2\pi K_R(g,p)}.
\label{appB:eta-from-stiffness}
\end{equation}
Therefore
\begin{equation}
\eta(g,p)
=
g^{-1/p}\eta(1,p)
=
\eta(1,p)
\left(\frac{T}{K}\right)^{1/p}.
\label{appB:eta-scaling}
\end{equation}
The exponent $1/p$ is fixed by the homogeneity of the noncompact
model, while the prefactor contains the finite renormalization
accumulated along the flow to the Gaussian fixed line.

Coarse graining therefore universalizes the spatial form of the
long-wavelength fluctuations, while the microscopic exponent $p$ survives
in the temperature dependence of the renormalized stiffness and hence in
$\eta(T)$.

Restoring compactness introduces vortices into this emergent
harmonic medium. Their far-field energy takes the logarithmic form
\begin{equation}
E_v
=
\pi K_R\ln L
+
E_{\mathrm{core}},
\label{appB:vortex-energy}
\end{equation}
providing the long-wavelength elastic basis for the BKT mechanism.
The bare vortex energy evaluated in a zero-temperature nonlinear
background is therefore not the relevant infrared free-energy cost: the
nontopological fluctuations must first be integrated out to determine the
elastic medium in which vortices interact.

\section{\dengrev{Monte Carlo observables and global finite-size scaling}}%
\label{sec:mc_fss}

\begingroup
\subsection{Monte Carlo data and quality criteria}

We summarize here the numerical measurements and the finite-size procedure
used for the phase boundary in Fig.~1(c) of the main text. Production runs were made on
periodic square lattices up to $L=2048$ using SW, ECMC, or mixed SW--ECMC
updates.  The sampling interval was selected from preliminary blocking tests.
For the present analysis a fine-window point is retained only when at least
$128$ final blocks remain and the largest residual direct-observable
correlation is below $0.2$.  These conditions are quality checks on a completed
run; they are not criteria for locating the transition.

Figure~\ref{fig:mc_chi_fine_all_p} collects the fine-window susceptibility data
for every simulated $p$, using the same curves as the numerical
data-processing record.  The finite-size evolution is appreciable even when
the individual statistical errors are small, so $T_c$ is determined from a
joint scaling analysis rather than from any single visual intersection.

\subsection{Global finite-size-scaling analysis}

For the BKT transition we use
$A_\chi(L,T)=\chi_0(L,T)/L^{7/4}$ and fit all retained sizes simultaneously to
the logarithmically corrected scaling form
\begin{align}
  Y_L(T)&\equiv
  A_\chi(L,T)(\ln L+C_1)^{-1/8}
  =a_0+a_1x+a_2x^2,
  \label{eq:global_fss_supp}\\
  x&=(T-T_c)(\ln L+C_2)^2 .
  \nonumber
\end{align}
The polynomial in $x$ is a local expansion of the smooth scaling function near
the transition.

Only $(T_c,C_1,C_2)$ are nonlinear parameters.  For every trial value of these
three parameters, the coefficients $(a_0,a_1,a_2)$ are obtained directly by
weighted linear least squares, after which the nonlinear objective is minimized
by a global search followed by bounded local refinement.  The statistical
weights are the blocking errors of the Monte Carlo points.  We repeat the
analysis for increasing $L_{\min}$.  \dengrevxxiv{A transition estimate is
regarded as controlled when $T_c$ remains within a common stability band and
the weighted goodness of fit is acceptable.  The primary analysis uses
$L_{\min}=64$.  Parametric bootstrap refits give the statistical component,
while the spread among stable fits gives the method component.  Fits that no
longer retain sufficient size coverage are used only as stability checks.}

\begin{figure}[t]
  \centering
  \includegraphics[width=\textwidth]{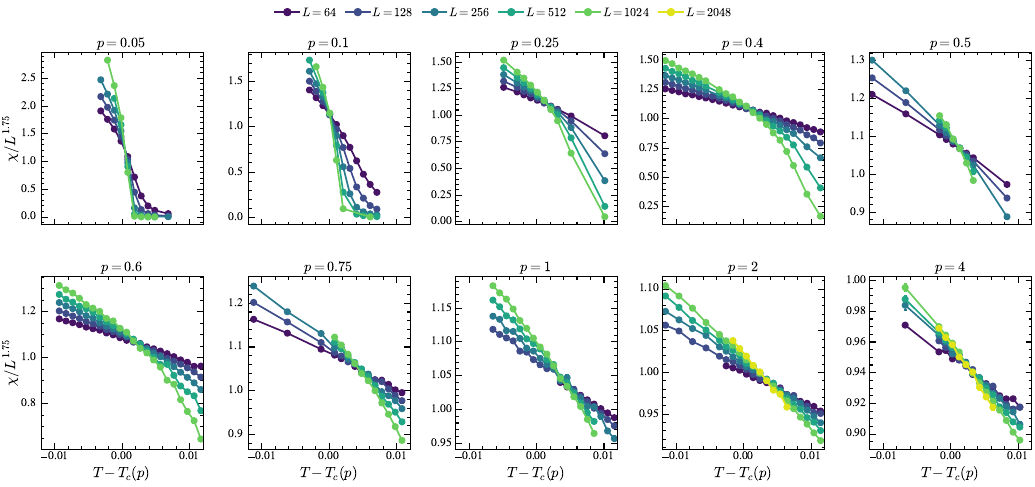}
  \caption{\label{fig:mc_chi_fine_all_p}%
    Fine-window susceptibility ratio $A_\chi=\chi_0/L^{7/4}$ for all
    simulated exponents versus the absolute temperature, so the transition
    temperature of each panel can be read directly against the
    $T_c$ values in Fig.~1(c) of the main text. The updated $p=1$ calibration, the
    $p=2$ extension to $T=0.620$, and the additional $p=2,4$ statistics
    through $L=2048$ are included.  Colors distinguish $L$; lines connect
    the measured temperatures in increasing order.}
\end{figure}

The global analysis is most constraining in the intermediate range, where the
near-critical data admit a common scaling function and the fitted $T_c$ is
stable against increasing $L_{\min}$.  At the smallest $p$, the fitted center
is comparatively stable but the low-order local form does not yet describe the
data at the level implied by their small pointwise errors.  At the largest $p$,
the dominant limitation is the weak temperature dependence and the balance of
near-critical coverage across sizes.  \dengrevxxiv{These cases are reported
from their stability envelopes rather than from the local covariance of a
single fit.  As a calibration, the same global analysis at $p=1$ gives
$T_c^{\mathrm{FSS}}=0.8933(9)$, whose uncertainty envelope contains the exact
standard-XY value $0.89294(8)$.  This agreement supports the uncertainty
envelopes quoted for the other exponents.}  The $\xi/L$ curves and the thermal collapse in
Fig.~2 of the main text remain independent consistency checks and are not included
as additional estimators of $T_c$.

\begin{table}[!h]
  \centering
  \footnotesize
  \caption{\label{tab:tc_crossing}Transition temperatures used in
  Fig.~1(c) of the main text. For controlled global fits, the quoted uncertainty
  \dengrevxxiv{combines the primary-fit bootstrap width and the spread among
  stable fits in quadrature.}  The smallest-$p$ centers are model limited and
  are therefore listed without a controlled one-standard-deviation interval.}
  \begin{ruledtabular}
  \begin{tabular}{ccp{0.47\columnwidth}}
    $p$ & $T_c$ & method / note \\
    \hline
    0.05 & $0.3071$ & model-limited center \\
    0.10 & $0.5057$ & model-limited center \\
    0.25 & $0.81124(41)$ & constrained global FSS \\
    0.40 & $0.9240(14)$ & constrained global FSS \\
    0.50 & $0.95204(68)$ & constrained global FSS \\
    0.60 & $0.9579(21)$ & constrained global FSS \\
    0.75 & $0.9446(12)$ & constrained global FSS \\
    1.00 & $0.89294(8)$ & \dengrevxxiv{exact benchmark; global FSS:
    $0.8933(9)$} \\
    2.00 & $0.6380(13)$ & global FSS; weak temperature sensitivity \\
    4.00 & $0.2845(30)$ & global FSS; weak temperature sensitivity \\
  \end{tabular}
  \end{ruledtabular}
\end{table}

\endgroup

\end{document}